\documentclass[aps,prl,reprint,superscriptaddress,amsmath,amssymb,amsfonts,showpacs]{revtex4-1}

\usepackage{graphicx}
\usepackage{dcolumn}
\usepackage{bm}
\usepackage{hyperref}
\usepackage{breakurl}
\usepackage{latexsym}
\usepackage{color}

\begin{document}


\title{Direct determination of the atomic mass difference of $^{187}$Re and $^{187}$Os for neutrino physics and cosmochronology}



\author{D.A. Nesterenko}
\affiliation{Max-Planck-Institut f\"ur Kernphysik, Saupfercheckweg 1, 69117 Heidelberg, Germany}
\affiliation{Petersburg Nuclear Physics Institute, Gatchina, 188300 St. Petersburg, Russia}

\author{S. Eliseev}
\affiliation{Max-Planck-Institut f\"ur Kernphysik, Saupfercheckweg 1, 69117 Heidelberg, Germany}

\author{K. Blaum}
\affiliation{Max-Planck-Institut f\"ur Kernphysik, Saupfercheckweg 1, 69117 Heidelberg, Germany}

\author{M. Block}
\affiliation{GSI Helmholtzzentrum f\"ur Schwerionenforschung GmbH, Planckstra{\ss}e 1, 64291 Darmstadt, Germany}

\author{S. Chenmarev}
\affiliation{Max-Planck-Institut f\"ur Kernphysik, Saupfercheckweg 1, 69117 Heidelberg, Germany}
\affiliation{Physics Faculty of St.Petersburg State University, 198904, Peterhof, Russia}

\author{A. D\"orr}
\affiliation{Max-Planck-Institut f\"ur Kernphysik, Saupfercheckweg 1, 69117 Heidelberg, Germany}

\author{C. Droese}
\affiliation{Institut f\"ur Physik, Ernst-Moritz-Arndt-Universit\"at, 17487 Greifswald, Germany}
\affiliation{Helmholtzinstitut Mainz, 55099 Mainz, Germany}

\author{P.E. Filianin}
\affiliation{Max-Planck-Institut f\"ur Kernphysik, Saupfercheckweg 1, 69117 Heidelberg, Germany}
\affiliation{Physics Faculty of St.Petersburg State University, 198904, Peterhof, Russia}

\author{M. Goncharov}
\affiliation{Max-Planck-Institut f\"ur Kernphysik, Saupfercheckweg 1, 69117 Heidelberg, Germany}

\author{E. Minaya Ramirez}
\affiliation{Max-Planck-Institut f\"ur Kernphysik, Saupfercheckweg 1, 69117 Heidelberg, Germany}

\author{Yu.N. Novikov}
\affiliation{Max-Planck-Institut f\"ur Kernphysik, Saupfercheckweg 1, 69117 Heidelberg, Germany}
\affiliation{Petersburg Nuclear Physics Institute, Gatchina, 188300 St. Petersburg, Russia}
\affiliation{Physics Faculty of St.Petersburg State University, 198904, Peterhof, Russia}

\author{L. Schweikhard}
\affiliation{Institut f\"ur Physik, Ernst-Moritz-Arndt-Universit\"at, 17487 Greifswald, Germany}

\author{V.V. Simon}
\affiliation{GSI Helmholtzzentrum f\"ur Schwerionenforschung GmbH, Planckstra{\ss}e 1, 64291 Darmstadt, Germany}
\affiliation{Helmholtzinstitut Mainz, 55099 Mainz, Germany}

\begin{abstract}
For the first time a direct determination of the atomic mass difference of $^{187}$Re and $^{187}$Os has been performed with the Penning-trap  mass spectrometer SHIPTRAP applying the novel phase-imaging ion-cyclotron-resonance technique. The obtained value of 2492(30$_{stat}$)(15$_{sys}$) eV is in excellent agreement with the $Q$-values determined indirectly with microcalorimetry and thus resolves a long-standing discrepancy with older proportional counter measurements. This is essential for the determination of the neutrino mass from the $\beta^-$-decay of $^{187}$Re as planned in the future microcalorimetric measurements. In addition, an accurate mass difference of $^{187}$Re and $^{187}$Os is also important for the assessment of $^{187}$Re for cosmochronology.  
\end{abstract}


\maketitle



The $\beta^-$-decay of $^{187}$Re plays a key role in neutrino physics, nuclear cosmochronology and the theory of extremely low-energy $\beta$-decay due to its remarkably small $Q$-value of about 2.5 keV.
This $\beta$-transition is considered one of the best candidates for the determination of the neutrino mass \citep{Arnaboldi-2003}. The MARE experiment based on cryogenic microcalorimetry plans to reach an uncertainty of 0.2 eV in the neutrino-mass determination after 10 years of data taking with a large array of 10$^5$ semiconductor thermistors \citep{MARE}. For the development of this experiment it is desirable to know the $Q$-value of the $\beta^-$-decay of $^{187}$Re with an accuracy of at least a few tens of eV.\\
The small $Q$-value of the $\beta^-$-decay of $^{187}$Re also implies a very long lifetime of 43.30(7) Gy for neutral $^{187}$Re atoms \citep{AME-2012}. This makes the pair $^{187}$Re-$^{187}$Os a suitable cosmic clock for the determination of the age of the Universe \citep{Clayton-1964}. The peculiarity of this cosmic clock is based on the fact that $^{187}$Re and $^{187}$Os are originally produced independently, in the rapid (r) and slow (s) neutron capture process, respectively. After production, $^{187}$Re decays into $^{187}$Os whose production mechanism in the s-process is well investigated \citep{Yokoi}. Knowing the present abundance of $^{187}$Os in nature, one can subtract the well known s-process component and determine the radiogenic contribution to the abundance of $^{187}$Os originated from the r-process. Thus, one can determine at what time in the past $^{187}$Re was produced and hence when the r-process took place. A possible additional production of $^{187}$Re from the s-process path through an isomer of $^{186}$Re only accounts for less than 1$\%$ contribution relative to the abundance of $^{186}$Os \citep{Hayakawa}. A thorough investigation of the s-process production by measuring the neutron capture and inelastic scattering cross-sections has been undertaken in \citep{Kaepp}. While, the $^{187}$Os abundance can be modified by the $\beta$-decay to a bound state in highly ionized $^{187}$Re ions, whose half-life can become as short as a few tens of years  \citep{Bosch-1996}, this modification is insignificant. Another origin for a possible modification of the $^{187}$Os abundance is the inverse transformation of $^{187}$Os to $^{187}$Re \citep{Arnold,Takahashi}. In hot stellar conditions, e.g., some of the low excited nuclear states of $^{187}$Os are in statistical equilibrium. They can participate in the electron capture process resulting in the production of $^{187}$Re. The strength of this process depends on astrophysical conditions and the energy balance between the ionic ground states of the isobaric nuclides, which can be derived from the mass difference of the neutral atoms. The directly measured mass difference can answer the question whether the electron capture from the excited states of $^{187}$Os in the stellar interior is energetically possible.\\
In addition, it is noteworthy that the small $Q$-value of the $\beta^-$-decay of $^{187}$Re allows a test of the existing $\beta$-decay theory in the regime of very small energy releases, when there is a strong influence of the atomic shell on the decay spectrum of very low energy electrons (so called screening effect), an effect of the non-pointlike charge of a nucleus and of the $\beta$-enviromental fine structure \citep{Benedek-1999}. For such a test, an accurate knowledge of the $Q$-value is essential.\\

\begin{figure} [t]
\includegraphics[width=0.48\textwidth]{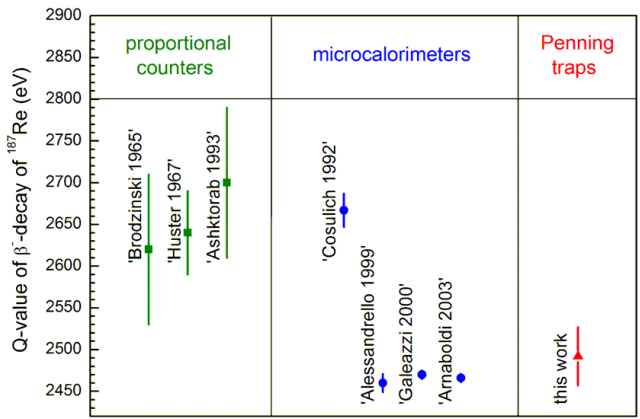}
\caption{\label{fig1} (color online) The $Q$-values of the $\beta^-$-decay of $^{187}$Re obtained in (Brodzinski-1965:\citep{Brodzinski-1965}, Huster-1967:\citep{Huster-1967}, Ashktorab-1993:\citep{Ashktorab-1993}, Cosulich-1992:\citep{Cosulich-1992}, Alessandrello-1999:\citep{Alessandrello-1999}, Galeazzi-2000:\citep{Galeazzi-2000}, Arnaboldi-2003:\citep{Arnaboldi-2003}) and in this work. The uncertainty of our value comprises the statistical and systematical uncertainties.}
\end{figure}

In this Letter we report on the first direct high-precision Penning-trap determination of the atomic mass difference of $^{187}$Re and $^{187}$Os. The experiment was performed with SHIPTRAP \citep{Block-2007} by a measurement of the cyclotron-frequency ratio of $^{187}$Re and $^{187}$Os ions, $R=\nu_c(^{187}Os^+)/\nu_c(^{187}Re^+)$. The cyclotron frequency $\nu_c$ of an ion with mass $m$ and charge $q$ in a magnetic field with strength $B$, given by $\nu_c=qB/(2\pi m)$, was determined as the sum of the two trap radial-motion frequencies: magnetron frequency $\nu_-$ and modified cyclotron frequency $\nu_+$, i.e., $\nu_c$=$\nu_-$+$\nu_+$.\\
Until now the $Q$-value of the $\beta^-$-decay of $^{187}$Re had only been determined indirectly as a fit parameter from the analysis of the $\beta^-$-decay spectrum. Fig.~\ref{fig1} shows two sets of data: One of which combines the values obtained with gas proportional counters \citep{Brodzinski-1965,Huster-1967,Ashktorab-1993} resulting in an average value of $Q$=2647(39) eV, whereas the other comprises the values obtained with cryogenic microcalorimetry \citep{Cosulich-1992,Alessandrello-1999,Galeazzi-2000,Arnaboldi-2003}. If the microcalorimetric result of \citep{Cosulich-1992}, which agrees with the proportional counter results, is ignored, then the average value of this group is $Q$=2466.6(1.6) eV. There is a substantial discrepancy between the $Q$-values given by different groups and methods. Thus, it is essential to perform an independent measurement of this $Q$-value with an uncertainty of at most a few tens of electron volts in order to resolve this discrepancy.\\
In particular, a significant deviation of the $Q$-value obtained by well established mass spectrometry from that obtained with cryogenic microcolorimetry would hint at the existence of systematic effects inherent in microcalorimetry, which would have a severe impact on the uncertainty of the planned experiments to determine the  neutrino mass with this technique.\\  
\begin{figure*} [tb]
\includegraphics[width=1\textwidth]{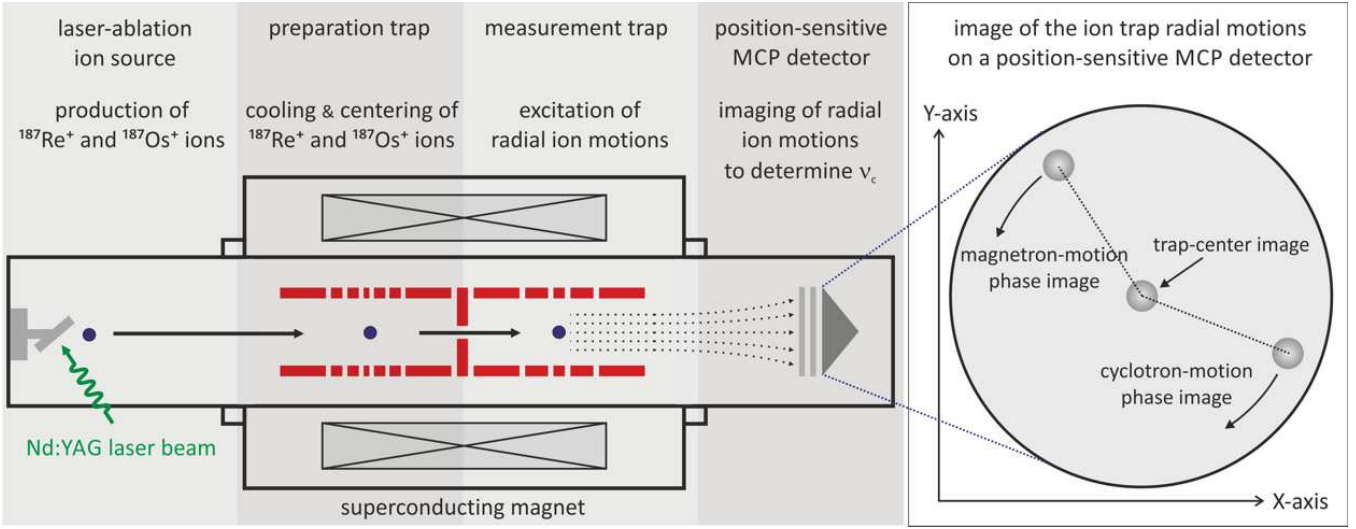}
\caption{\label{fig2} (color online) Schematic of the SHIPTRAP setup used for the determination of the $Q$-value of the $\beta^-$-decay of $^{187}$Re. Note that while the ions perform cyclotron and magnetron revolutions in the same sense, their cyclotron phase image is inverted during the cyclotron-to-magnetron conversion \citep{PI-ICR-2}.}
\end{figure*}\\         
A schematic of the experimental setup is presented in Fig.~\ref{fig2}.
In our experiment singly-charged ions of $^{187}$Re and $^{187}$Os were produced with a laser-ablation ion source \citep{Laser} by irradiating the corresponding metallic samples of natural Re and Os with a frequency-doubled Nd:YAG laser beam. The ions were transferred from the source into a preparation trap (PT). There, the ions were cooled and centered via mass-selective buffer-gas cooling \citep{Cooling}. Afterwards, their cyclotron frequencies were measured in a measurement trap (MT) with the novel Phase-Imaging Ion-Cyclotron Resonance technique (PI-ICR) \citep{PI-ICR-1,PI-ICR-2}. $^{187}$Re$^+$ and $^{187}$Os$^+$ cyclotron frequencies were measured alternately.\\
The cyclotron frequency $\nu_c$ of the corresponding nuclide was measured directly by applying measurement scheme 2 described in detail in \citep{PI-ICR-2}. This measurement scheme is as follows (see \citep{PI-ICR-2} for details): After cooling and centering the ions of interest in the PT, the ions are transferred into the center of the MT. Then, the coherent components of the magnetron and the axial motions are damped via dipole rf-pulses at the corresponding motion frequencies. After this preparatory step, the radius of the ion cyclotron motion is increased to a certain radius in order to set the initial phase of the cyclotron motion. Then, two excitation patterns are applied alternately in order to measure the ion cyclotron frequency $\nu_c$. In pattern 1 the cyclotron motion is first converted to the magnetron motion with the same radius. Then, the ions perform the magnetron motion for the time $t$ accumulating a certain magnetron phase. After time $t$ has elapsed, the ions' position in the trap is projected onto a position-sensitive detector by ejecting the ions from the trap towards the detector \citep{Eitel}. In pattern 2 the ions first perform the cyclotron motion for the time $t$ accumulating a certain cyclotron phase with a consecutive  conversion to the magnetron motion and again projection of the ion position in the trap onto a position-sensitive detector. The angle between the ion-position images corresponding pattern 1 and 2, respectively, with respect to the trap center image is proportional to the ion cyclotron frequency $\nu_c$.\\        
Patterns 1 and 2 are called in this work "magnetron-motion phase" and "cyclotron-motion phase", respectively, because during the corresponding pattern the ions mostly perform a magnetron or cyclotron motion, respectively.  Pulse patterns 1 and 2 were applied for a total measurement time of approximately 5 minutes. The duration of each pulse pattern was about 700 ms. Thus, on the measurement scale of 5 minutes the "magnetron-motion phase" and "cyclotron-motion phase" can be considered to be measured simultaneously. After injecting the ions into the MT and before measuring the corresponding trap-motional phase over 700 ms, the coherent components of the magnetron and axial motions were damped to amplitudes of about 0.01 mm and 0.4 mm, respectively, by applying 1 ms dipole rf-pulses at the corresponding motional frequencies. These steps are required to reduce the shift in the ratio of the $^{187}$Os$^+$ and $^{187}$Re$^+$ ions due to the anharmonicity of the trap potential, inhomogeneity of the magnetic field and conversion of the cyclotron motion to magnetron motion to a level well below 10$^{-10}$ (see \citep{PI-ICR-2} for details). After damping, the cyclotron motion was excited to an amplitude of about 0.5 mm by a 1 ms dipole rf-pulse at the modified cyclotron frequency of the corresponding nuclide. Then, the ions accumulated the cyclotron-motion phase over 700 ms for the measurement of the "cyclotron-motion phase", before being ejected toward the position-sensitive MCP-detector. For the measurement of the "magnetron-motion phase" the cyclotron motion was immediately converted to magnetron motion with subsequent accumulation of the magnetron phase for 700 ms.\\
Data with more than 5 detected ions per cycle were not considered in the analysis in order
to reduce a possible cyclotron frequency shift due to ion-ion interaction. In Fig.~\ref{fig3} a 5 min.-measurement of the cyclotron frequency $\nu_c$ of $^{187}$Os$^+$ ions is presented. The positions of the magnetron-motion and cyclotron-motion phase spots were chosen such that the angle $\alpha_c=\alpha_{mag}-\alpha_{cyc}$ between the phase spots calculated with respect to the center of the MT did not exceed few degrees. This was required to reduce the shift in the ratio of the $^{187}$Os$^+$ and $^{187}$Re$^+$ ions due to the conversion of the cyclotron motion to magnetron motion and the possible distortion of the ion-motion projection onto the detector to a level well below 10$^{-10}$ \citep{PI-ICR-2}. The relation between $\alpha_c$ and the cyclotron frequency $\nu_c$ is given by\\
\begin{equation}
	\nu_c=(\alpha_c + 2 \pi n)/2 \pi t,
\end{equation}
where $n$ is the number of revolutions the investigated ion would perform in a pure magnetic field $B$ during the phase-accumulation time $t$.   
\begin{figure} [tb]
\includegraphics[width=0.48\textwidth]{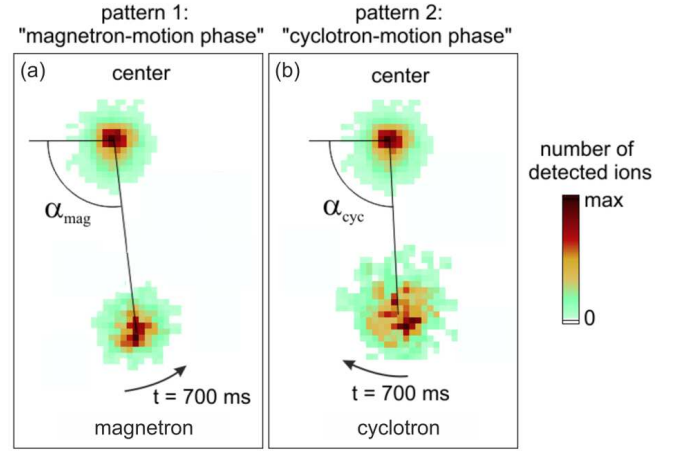}
\caption{\label{fig3} (color online) XY-distributions of $^{187}$Os$^+$ ions on the position-sensitive MCP detector. There are the central spot and the phase spots corresponding to the ions in the center of trap, ions after applying pattern 1 of excitation-pulse scheme for accumulating the magnetron-motion phase (a)  and pattern 2 for accumulating the cyclotron-motion phase (b), respectively \citep{PI-ICR-2}. For details see text.}
\end{figure}\\         
The cyclotron frequencies $\nu_c$ of the $^{187}$Os$^+$ and $^{187}$Re$^+$ ions were measured alternately for several days. For the single ratio of the cyclotron frequencies $\nu_c$ of the $^{187}$Os$^+$ and $^{187}$Re$^+$ ions determined at the time $t$, the cyclotron frequency of nuclide 1 measured right before and after the cyclotron frequency of nuclide 2 was linearly interpolated to the measurement time of the cyclotron frequency of nuclide 2. For each of the 33 4-hour periods the weighted mean ratio $R_{4 hour}$ of the single ratios was calculated along with the inner and outer errors \citep{Birge}. The final cyclotron-frequency ratio $R$ is the weighted mean of the $R_{4 hour}$ ratios, where the maximum of the inner and outer errors of the $R_{4 hour}$ ratios were taken as the weights to calculate $R$. The difference between the inner and outer errors does not exceed 10$\%$.\\
In Fig.~\ref{fig4} the $Q$-values of the $\beta^-$-decay of $^{187}$Re calculated from the cyclotron-frequency ratios $R_{4 hour}$ are shown for the entire measurement period. The final frequency ratio $R$ with its statistical and systematic uncertainties as well as the corresponding $Q$-value are $R$=1.000$\,$000$\,$014$\,$31(17)(9) and $Q$=2492(30)(15) eV, respectively. The systematic uncertainty in the frequency-ratio determination originates from the anharmonicity of the trap potential, the inhomogeneity of the magnetic field, the distortion of the ion-motion projection onto the detector, and the conversion of the cyclotron motion to the magnetron motion \citep{PI-ICR-2}.\\
In addition, the frequency ratio $R_{Xe}=\nu_c(^{131}$Xe$^+)/\nu_c(^{132}$Xe$^+)$ of the two stable isotopes of xenon - $^{131}$Xe and $^{132}$Xe - was measured in a similar manner and the mass difference $\Delta M_{Xe}$ was calculated yielding $R_{Xe}$=1.007$\,$632$\,$057$\,$62(20)(12) and $\Delta M_{Xe}$=930628611(25)(15) eV.  $\Delta M_{Xe}$ obtained in our experiment is in excellent agreement with the mass difference recently determined with the Penning-trap mass spectrometer FSU-trap ($M(^{132}$Xe)-$M(^{131}$Xe)=930628604(13) eV) differing by 7(32) eV \citep{FSU}. This provides an additional cross check for the accuracy of our measurement. Note that unlike for the mass doublet $^{187}$Re-$^{187}$Os, the mass difference of two xenon isotopes is one mass unit. Thus, for the first time the mass difference of \textit{singly charged non mass-doublets} was measured with a relative uncertainty of 0.2 ppb. A detailed analysis of the Xe data will be published later.\\
\begin{figure} [htb]
\includegraphics[width=0.48\textwidth]{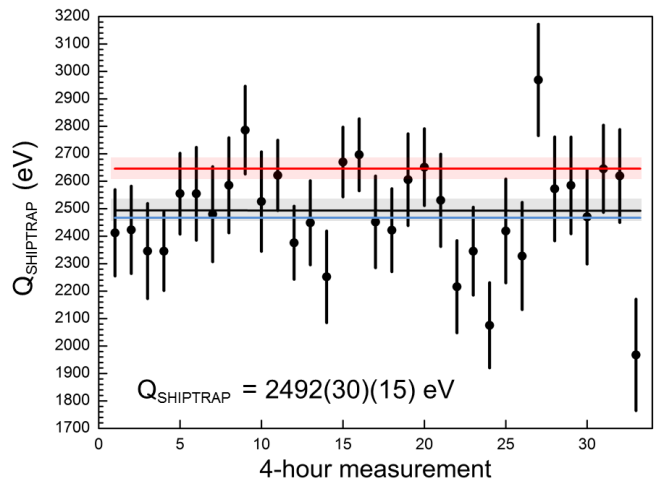}
\caption{\label{fig4} (color online) The $Q$-values of the $\beta^-$-decay of $^{187}$Re calculated from the cyclotron-frequency ratios $R_{4 hour}$. The black line and the grey shaded band are the average $Q$-value and its uncertainty of the work reported here. The red line and the red shaded band are the average $Q$-value and its uncertainty, respectively, obtained with proportional counters. The blue line represents the average $Q$-value obtained with cryogenic microcalorimeters. The thickness of the line exceeds the uncertainty of this $Q$-value.}
\end{figure}\\

Our result for the atomic mass difference of $^{187}$Re and $^{187}$Os is in perfect agreement with the latest microcalorimetric measurements: $Q$ = 2460(11) eV \citep{Alessandrello-1999}, $Q$ = 2470(4) eV \citep{Galeazzi-2000} and $Q$ = 2466.1(1.7) eV \citep{Arnaboldi-2003} with an average value of $Q$ = 2466.6(1.6) eV (see Fig.~\ref{fig1}). Thus, on the level of the present accuracy there are no unexpected systematic effects inherent in the cryogenic microcalorimetric technique. For the determination of the neutrino mass the $Q$-value must be determined with a substantially lower uncertainty - on the sub-eV level. At present, there are no Penning-trap experiments capable of performing such precise $Q$-value measurements. This will become possible with the realization of the PENTATRAP experiment  \citep{Repp,Roux}.\\
In addition, the measured mass difference allows an assessment of the probability of electron capture in ionic $^{187}$Os in hot stellar conditions. The strength of this process depends on the relative thermal population $f$ of the nuclear excited states, which in turn depends on the energy $E^*$ and spin values $I^*$ and $I$ of the excited and ground states of $^{187}$Os, respectively, and the temperature $T$ of the environment: $f=(2I^*+1)/(2I+1) \times exp((-E^*)/kT)$ \citep{Cameron-1959}. It also depends on the energy of the transition, which contains the mass difference of the neutral ground states.  The electron-capture process takes place if
\begin{equation}
\begin{split}
	\Delta_{i}=M(^{187}\text{Os}^{q+})+E^*-B_i-M(^{187}\text{Re}^{q+})=\\
	E^*-B_i-Q+(B(\text{Os})-B(\text{Re}))-\\
	(B^{76-q}(\text{Os})-B^{75-q}(\text{Re}))>0,
\end{split}
\end{equation}    
where $B_i$ is the modulus of the binding energy of the captured electron, $M(^{187}$Re$^{q+})$ and $M(^{187}$Os$^{q+})$ are the masses of $^{187}$Re and $^{187}$Os ions in the charge state of $q+$, respectively, $B($Os$)-B($Re$)$ is the modulus of the difference of the total binding energy of electrons in $^{187}$Os and $^{187}$Re atoms, and $B^{76-q}($Os$)$ and $B^{75-q}($Re$)$ are the modulus of the binding energies of (76-q) and (75-q) electrons in $^{187}$Os$^{q+}$ and $^{187}$Re$^{q+}$ ions, respectively. The nuclear recoil energy can be neglected. We also assumed a smallness of the ionic excitation energy differences \citep{Johnson}. Osmium nuclide has a few nuclear excited states which can be thermally populated in the hot stellar conditions. For example, the first excited nuclear state of $^{187}$Os at an energy of 9756(19) eV \citep{NDS-2009} has a thermal population of about 7 $\%$  at $T\approx 3.5\times10^7$ K. As an example, for this excitation energy and temperature, we consider the partial M$_1$-capture from the 9756(19) eV excited state by the Os$^{63+}$ ions: The $\Delta_{i}$-value for this electron capture of about 2.7 keV (the binding energies of the electrons are taken from \citep{Rodriguez-2004,Larkins-1977}) allows the electron capture in $^{187}$Os. Another excited level in $^{187}$Os with an energy of 190.56 keV can be quite sizably (2$\%$) populated at a rather high temperature of 5.8$\times$10$^8$ K. For this level the $\Delta_{K}$ is estimated to be 42.9 keV. For a typically assumed s-process temperature of 3.5$\times$10$^8$ K the levels with energies of 74.3, 75.0 and 100.4 keV can also contribute to the stellar capture process.   This opens an alternative production process of $^{187}$Re other than the r-process. Thus, it is essential to know all possible chains of mutual transformations between $^{187}$Re and $^{187}$Os for a correct use of $^{187}$Re/$^{187}$Os as a cosmic clock. Here, we just point to one of such transformations, the reverse decay of $^{187}$Os to $^{187}$Re. A more detailed consideration of this issue is beyond the scope of this work and will be presented elsewhere.\\

In summary, the atomic mass difference of $^{187}$Re and $^{187}$Os has been determined with the Penning-trap mass spectrometer SHIPTRAP with the novel PI-ICR technique. The measurement has yielded the value of 2492(30)(15) eV in a perfect agreement with the latest $Q$-values obtained with cryogenic microcalorimetry thus solving the puzzle on the conflicting $Q$-values obtained by different groups and methods. In addition it was shown that a possibility of electron capture by $^{187}$Os ions in hot stellar conditions has to be considered for cosmochronology.\\

This work was supported by the Max-Planck Society, by the EU (ERC grant number 290870-MEFUCO), by the German BMBF (05P12HGFN5 and 05P12HGFNE) and by the Russian Ministry of Education and Science (project 2.2). We would like to thank V.P. Chechev and V. Shabaev for discussions.

\bibliography{Re}

\end{document}